\documentclass{article}
\usepackage{float}
\usepackage{comment}
\usepackage{graphicx}
\usepackage[numbers]{natbib}

\providecommand{\keywords}[1]{\textbf{Keywords:} #1}

\makeatletter
\renewenvironment{abstract}{%
    \if@twocolumn
      \section*{\abstractname}%
    \else 
      \begin{center}%
        {\bfseries \large\abstractname\vspace{\z@}}
      \end{center}%
      \quotation
    \fi}
    {\if@twocolumn\else\endquotation\fi}
\makeatother

\begin{document}


\title{Autocorrelation of returns in major cryptocurrency markets}

\author{%
\begin{tabular}{c} 
\textbf{Eugene Tartakovsky}$^1$ \\ eugene.t@3jane.com \\ \\
\textbf{Ksenia Plesovskikh}$^3$ \\ ksenia.p@rcdb.io 
\end{tabular} 
\and
\begin{tabular}{c} 
\textbf{Anastasiia Sarmakeeva}$^2$ \\ asarmakeeva@gmail.com \\ \\
\textbf{Alexander Bibik}$^3$ \\ alex.b@rcdb.io \\ 
\end{tabular} 
}

\date{\normalsize%
    $^1$3Jane Capital\\%
    $^2$Institute of Mechanics, Ural Branch of Russian Academy of Sciences\\%
    $^3$Research Center for Digital Assets and Blockchain\\[2ex]%
    \large\today
}

\maketitle

\begin{abstract}

This paper is the first of a series of short articles that explore the efficiency of major cryptocurrency markets.
A number of statistical tests and properties of statistical distributions will be used to assess if cryptocurrency markets are efficient, and how their efficiency changes over time.
In this paper, we analyze autocorrelation of returns in major cryptocurrency markets using the following methods: 
Pearson's autocorrelation coefficient of different orders, Ljung-Box test, and first-order Pearson's autocorrelation coefficient in a rolling window. All experiments are conducted on the BTC/USD, ETH/USD, ETH/BTC markets on Bitfinex exchange, and the XBT/USD market on Bitmex exchange, each on 5-minute, 1-hour, 1-day, and 1-week time frames.
The results are represented visually on charts. Statistically significant autocorrelation is persistently present on the 5m and 1H time frames on all markets. The tests disagree on the 1D and 1W time frames.
The results of this article are fully reproducible. Used datasets, source code, and a runnable Jupyter Notebook are available on GitHub\footnote{https://github.com/3jane/articles/tree/master/1-autocorrelation-time-bars}.

\end{abstract}


\section{Introduction}
In this paper, we use benchmarks provided by the efficient market hypothesis (EMH) \cite{the_efficient_market_on_the_stock_market} to estimate the efficiency of cryptocurrency markets. 

The cryptocurrency market is a new and fast-growing sector that adopts features from the financial markets established earlier. It has both advantages and disadvantages in comparison to the traditional markets. The main advantages are worldwide 24/7/365 trading and fast (often immediate) execution and settlement. 

The prevalent disadvantages are associated with the immaturity of the cryptocurrency markets. There are not enough established tools, more services and businesses are needed to engender competition and boost reliability, and the cryptocurrency market does not yet interface comfortably with the established traditional markets. 
For example, there were no prime brokers, and until recently, no custodians that provided services for cryptocurrency markets.

The market also has completely novel elements because the currencies being traded are issued on public blockchain technology, as described by Nakamoto in the original Bitcoin technical paper, \cite{Nakamoto}
and are therefore not controlled by any government or institution. All transactions are immutable, visible to everyone, and remain stored as long as the underlying blockchain exists.

The Efficient Market Hypothesis has been the central proposition of finance since the early 1970s, and it is one of the most controversial and well-studied propositions in all the social and economic sciences.
It states that asset prices reflect all available information as Malkiel's and Fama's research showed in \cite{Fama}, and that profiting from predicting price movements is very difficult and unlikely as Clarke et al. state in \cite{Clarke}. 

Over time, the attempts to prove or refute the EMH have been inconsistent, and we still cannot reliably claim that the EMH describes financial markets well enough to make investment decisions.
However, it provides us with a convenient unified test suite to detect when a market is definitely not efficient. 
That information can be used to direct further research on how to use the found inefficiency profitably .
By now, the majority of established financial markets have been analyzed by many researchers. 
Among others, Sewell in \cite{Sewell} explores the market efficiency of the Dow Jones Industrial Average (DJIA), and Zeman et al. \cite{Forex} do the same for EUR/USD foreign exchange market.

\section{State-of-the-art}

The Efficient Market Hypothesis is defined as follows. A market is said to be {\it efficient} with respect to an information set if the price would be unaffected by revealing the information set to all market participants, as the Malkiel research showed \cite{Malkiel}, i.e., if the price fully reflects that information set \cite{Fama}. The classic taxonomy of information sets, according to Roberts \cite{Roberts} and Malkiel and Fama \cite{Fama}, consists of the following:
\begin{itemize}
    \item {\it Weak form of efficiency}. The information set only includes the history of prices.
    \item {\it Semi-strong form of efficiency}. The information set includes all the information known to all market participants (publicly available information).
    \item {\it Strong form of efficiency}. The information set includes all the information known to any market participant (private information).
\end{itemize}

There exists some controversy over the efficiency of the cryptocurrency market based on the proponents and opponents of the EMH. For example, Nadarajah and Chu \cite{Nadarajah},  Bariviera \cite{Bariviera}, and  Tiwari et al. \cite{Tiwari} conclude that the Bitcoin market is almost efficient. In contrast, Jiang et al. \cite{Yonghong}, Chean et al. \cite{Cheah}, and Al-Yahyaee et al. \cite{Yahyaee} present skeptical empirical results that do not support the EMH for this market \cite{skeptical}.

Dr. Andrew Urquhart in his published work in 2016 \cite{Urquhart} examined the Bitcoin Market during that period and concluded that the market was not efficient, but was moving towards efficiency. In his work, he investigated the behavior of returns using the Ljung-Box test, runs tests, the dispersion ratio test, and the BDS test. The analysis showed that the Bitcoin market did not exhibit weak efficiency during the entire sampling period, but several market efficiency tests show that the market could become more efficient over time, suggesting that Bitcoin's yield was random later in the period. However, Bitcoin's inefficiency was quite strong. Three years have passed since that article was written and the situation in the Bitcoin market has changed significantly. Hence, we revisit the analysis to bring the information up to date.

One of the most recent publications on this topic by Ladislav Kristoufek and Miloslav Vošvrda \cite{Ladislav} is devoted to testing whether the examined Bitcoin, DASH, Litecoin, Monero, Ripple and Stellar were efficient, and the overall level of efficiency in the cryptocurrency market. The authors utilized the Efficiency Index comprising long-range dependence, fractal dimension, entropy components, and descriptive statistics. They came to the following conclusions:
\begin{itemize}
    \item Historically, every cryptocurrency was inefficient over the analyzed period.
    \item Efficiency and ranking were dependent on the quote currency (US dollar or Bitcoin).
    \item Most of the coins and tokens were efficient between July 2017 and June 2018.
    \item The least efficient coins were Ethereum and Litecoin, while DASH was found to be the most efficient cryptocurrency.
\end{itemize}

Another recent paper by Akihiko Noda \cite{skeptical} studies whether the market efficiency of major cryptocurrencies changes over time based on the adaptive market hypothesis (AMH). Empirical results in this work have shown that the degree of market efficiency changes over time, that the level of efficiency of the Bitcoin market is higher than other markets during most periods, and the efficiency of the cryptocurrency market has improved. Adaptive Market Hypothesis (AMH) embraces EMH as an idealization that is economically unrealizable but which serves as a useful benchmark for measuring relative efficiency. 


Presently, even with the cryptocurrency market taking root and occupying a prominent position in global finance, there is little research on it as a full-fledged player, hence the relevance of present study. First, we elaborate on how the statistics were calculated, then provide an analysis for the last 5 years, from 2014-07-01 to 2019-07-01. In conclusion, we evaluate and discuss the obtained results.

\section{Research hypothesis}
Consider the BTC/USD, ETH/USD, ETH/BTC markets on Bitfinex, and the XBT/USD market on Bitmex. Let us assume the markets to be efficient as a null hypothesis. For a market to be efficient according to EMH:

\begin{itemize}

\item There should be no statistically significant autocorrelation of any order on any time frame.

\item The p-value of Ljung-Box test should not drop below 0.05 on any lag and any time frame.

\item Rolling autocorrelation should display the same stable, close to zero results on any time frame.

\end{itemize}

\section{Research methodology}

\subsection{Pearson's autocorrelation coefficient}
A necessary but not sufficient condition for EMH to hold is that time series has no autocorrelation of any order. Let $X$ be a stochastic process and $t$ a point in time, then $X_t$ is the value produced by a given run of the process at time $t$. Suppose that $X_t$ has mean $\mu$ and variance $\sigma^2$ at time $t$ for each $t$. Then the Pearson autocorrelation coefficient between times $t_1$ and $t_2$ is defined by:
$$
R(t_1,t_2) = \frac{E[(X_{t_1} - \mu_{t_1})(X_{t_2} - \mu_{t_2})]}{\sigma_{t_1}\sigma_{t_2}},
$$
where $E$ -- the expected value operator.

\subsection{Ljung-Box test}

The null hypothesis of the Ljung-Box test \cite{Ljung-Box}, where $H_0$ signifies that the data are independently distributed (i.e., the correlations in the sample are taken as 0, so that any observed correlations in the data result from the randomness of the sampling process). The alternate hypothesis, $H_a$ -- the data are not independently distributed; they exhibit serial correlation.

A significant p-value in this test rejects the null hypothesis that the time series is not autocorrelated. To run the Ljung-Box test, we calculate the statistic $Q$. For a time series of length $n$:
$$
Q(h) = n(n + 2)\sum^h_{j=1}\frac{\hat{\rho}^2_k}{n - k}.
$$
Where $\hat{\rho}^2_k$ -- is the sample autocorrelation at lag $k$, and $h$ is the number of lags being tested. Under $H_{0}$, the statistic $Q$ asymptotically follows a $\chi _{{(h)}}^{2}$. For significance level $\alpha$, the critical region for rejection of the hypothesis of randomness is:
$$
Q > \chi_{1-\alpha,h}^2,
$$
where $\chi_{1-\alpha,h}^2$ is the $1-\alpha$ -- quantile of the chi-squared distribution with $h$ degrees of freedom.

\subsection{Pearson's autocorrelation coefficient in a rolling window}
Additionally, we compute first-order Pearson's autocorrelation coefficient in a rolling 1-year window. That will allow us to see how the coefficient changes over time and whether there is a stable deviation from zero. If there is, that would contradict the null hypothesis as well.

\section{Results}

All calculations were made using Python programming language, statsmodels \cite{1}, and pandas \cite{2} libraries. The four currency pairs chosen to carry out this research were selected for the following reasons:
\begin{itemize}
    \item BTC/USD on Bitfinex exchange: one of the oldest and most liquid spot cryptocurrency markets.
    \item XBT/USD perpetual swap on Bitmex futures exchange: the most liquid cryptocurrency market in the world (as of the date this paper was written) with 2.20 billion USD traded daily, according to their reports. 
    \item ETH/USD on Bitfinex exchange: the second-biggest cryptocurrency and one of the most liquid spot markets.
    \item ETH/BTC on Bitfinex exchange: a cross-pair between the first and second-most liquid cryptocurrencies. This pair exhibits significantly different behavior compared to USD-denominated pairs.
\end{itemize}

For each market, we conducted all tests using four different time frames: 5-minute, 1-hour, 1-day, and 1-week bars.

\subsection{Pearson's autocorrelation coefficient}

\begin{figure}[H]
    \centering
    \includegraphics[scale=0.2]{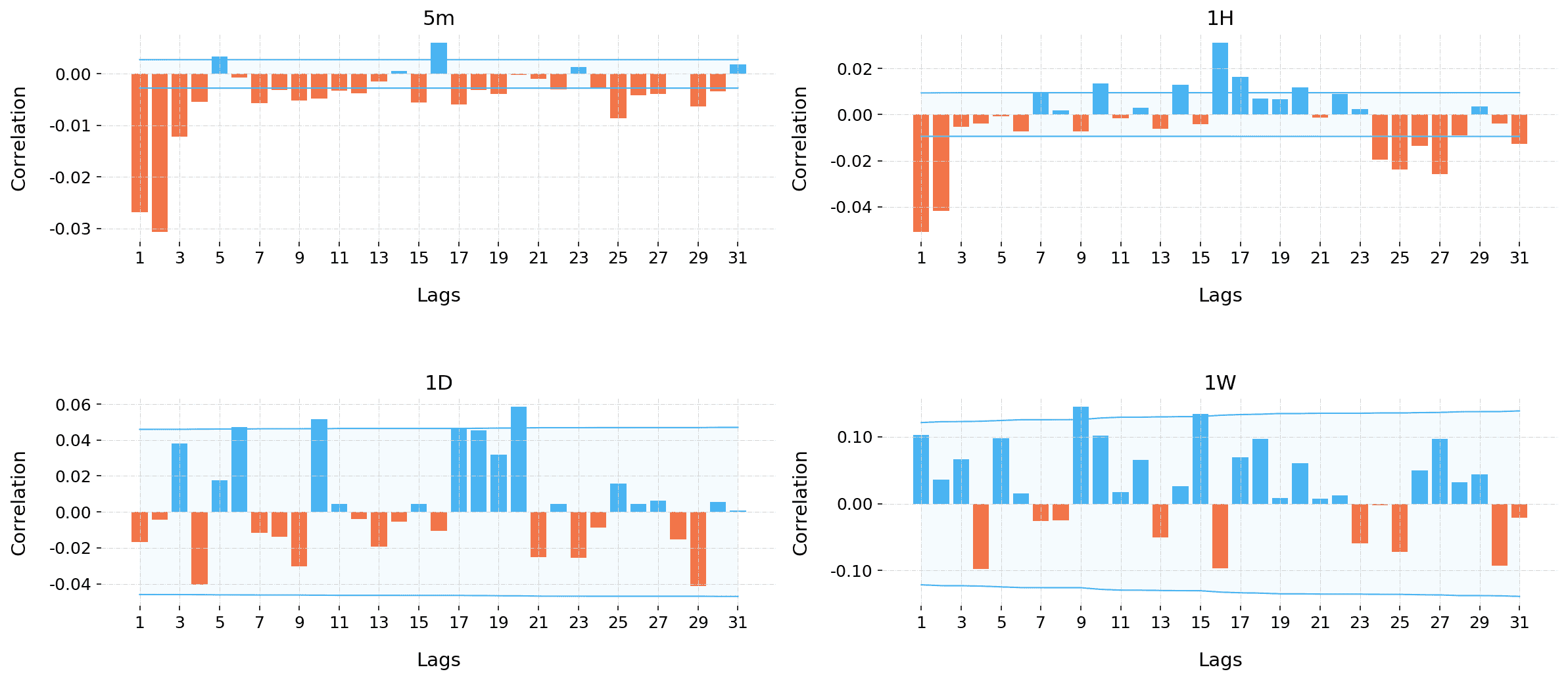}
    \caption{Autocorrelation of BTC/USD, from 2014-07-01 to 2019-07-01}
    \label{fig:btcusd_acorr}
\end{figure}

\begin{figure}[H]
    \centering
    \includegraphics[scale=0.2]{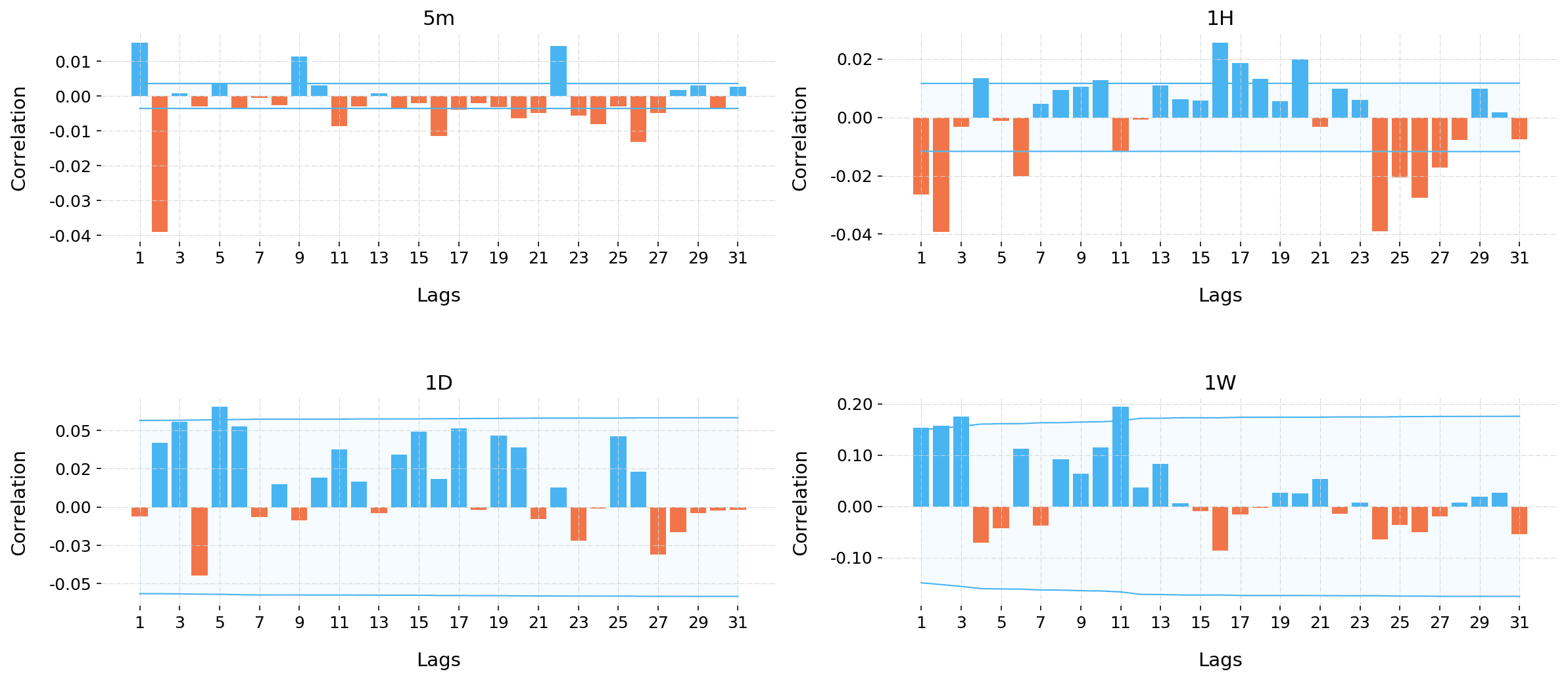}
    \caption{Autocorrelation of ETH/USD, from 2016-03-09 to 2019-07-01}
    \label{fig:ethusd_acorr}
\end{figure}

\begin{figure}[H]
    \centering
    \includegraphics[scale=0.2]{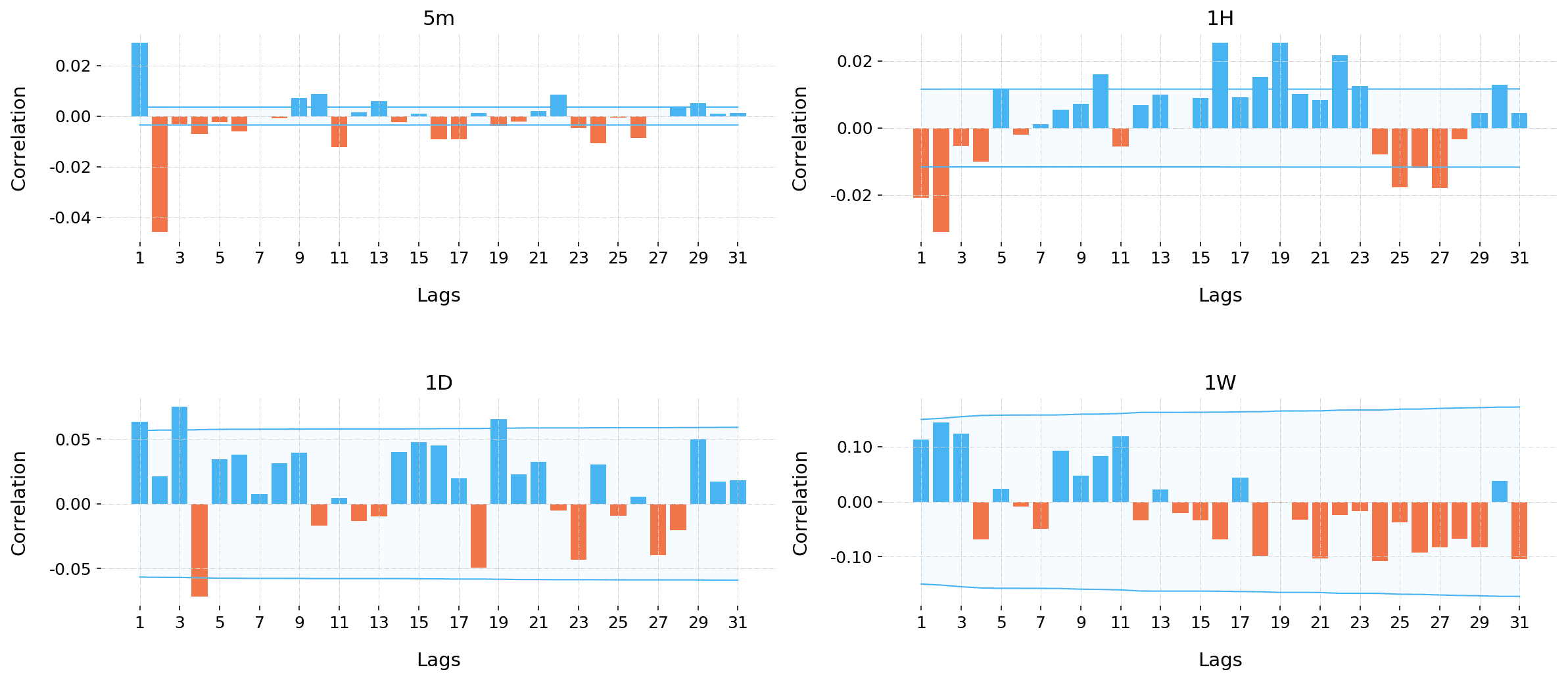}
    \caption{Autocorrelation of ETH/BTC, from 2016-03-09 to 2019-07-01}
    \label{fig:ethbtc_acorr}
\end{figure}

\begin{figure}[H]
    \centering
    \includegraphics[scale=0.2]{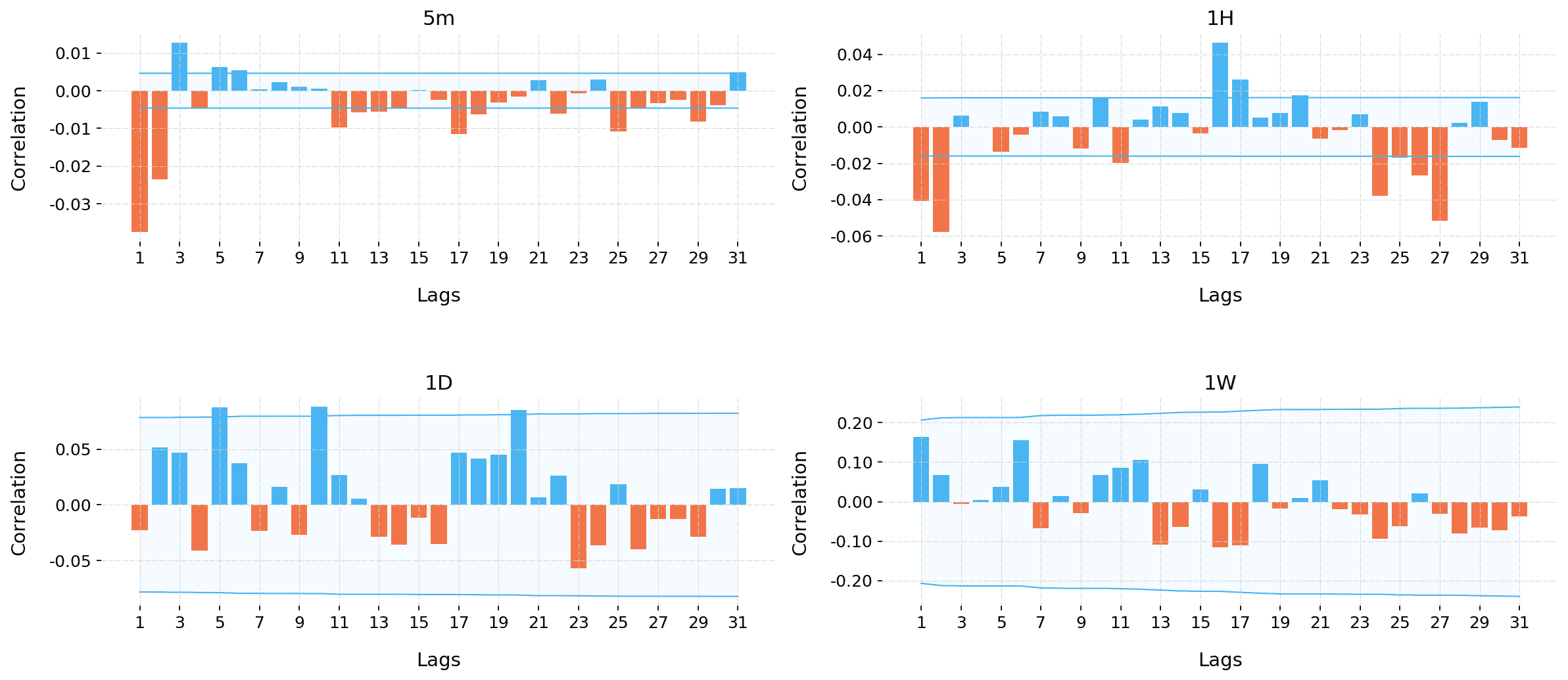}
    \caption{Autocorrelation of XBT/USD, from 2017-10-12 to 2019-07-01}
    \label{fig:xbtusd_acorr}
\end{figure}

Pearson autocorrelation coefficient was measured for the BTC/USD, ETH/USD, ETH/BTC markets on Bitfinex exchange and the XBT/USD market on Bitmex exchange, each on 5m, 1H, 1D and 1W time frames. Results are presented in figures  \ref{fig:btcusd_acorr} to \ref{fig:xbtusd_acorr}.

As shown in figures \ref{fig:btcusd_acorr} and \ref{fig:xbtusd_acorr}, statistically significant lower-order negative autocorrelation was found in the BTC/USD and XBT/USD markets for the 5m and 1H time frames. In contrast, on both ETH pairs in figures \ref{fig:ethusd_acorr} and \ref{fig:ethbtc_acorr}, first-order autocorrelation was found to be positive for the 5m and 1H time frames. Second-order autocorrelation was found to be negative. For all four markets, we observed statistically significant positive autocorrelation on the 16th lag of the 1H time frame. For the 1D and 1W time frames, there is no autocorrelation level of any order on any market that significantly exceeds the confidence interval. 

According to our observations, none of the four explored markets on lower time frames satisfy the no-autocorrelation condition required for a market to be considered efficient.

\subsection{Ljung-Box test}

\begin{figure}[H]
    \centering
    \includegraphics[scale=0.2]{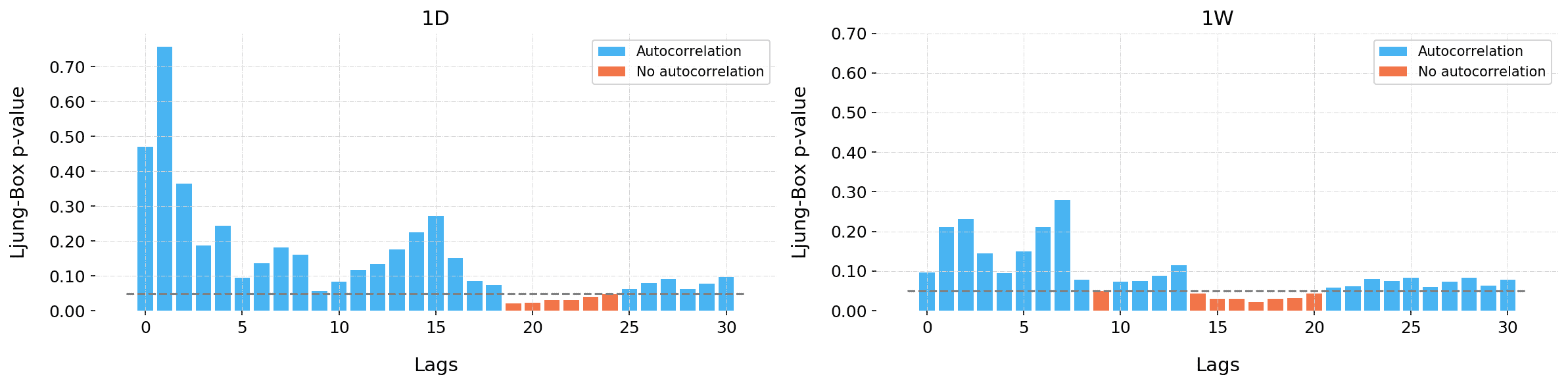}
    \caption{Ljung-Box test on BTC/USD, from 2014-07-01 to 2019-07-01}
    \label{fig:btcusd_lbtest}
\end{figure}

\begin{figure}[H]
    \centering
    \includegraphics[scale=0.2]{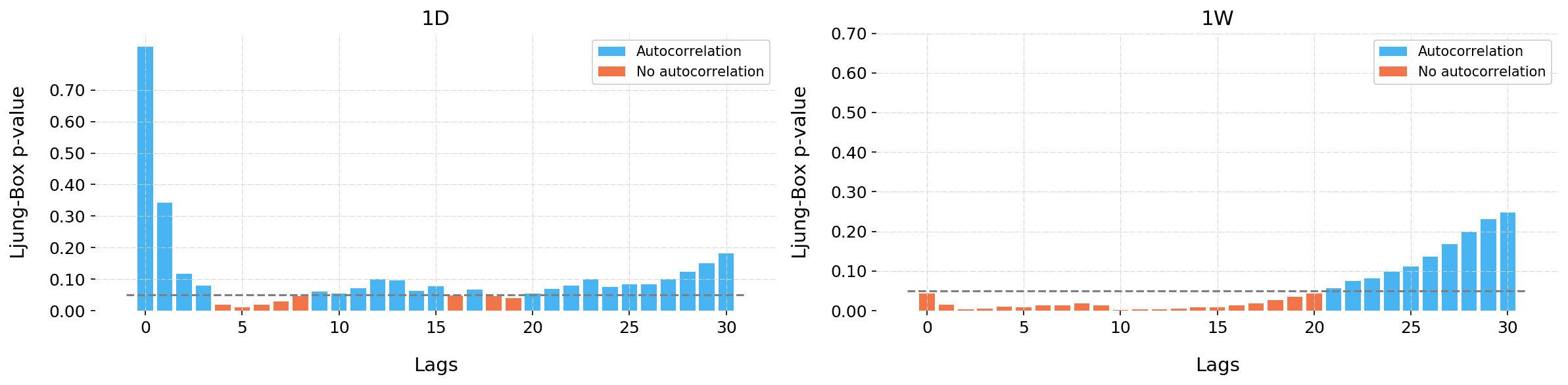}
    \caption{Ljung-Box test on ETH/USD, from 2016-03-09 to 2019-07-01}
    \label{fig:ethusd_lbtest}
\end{figure}

\begin{figure}[H]
    \centering
    \includegraphics[scale=0.2]{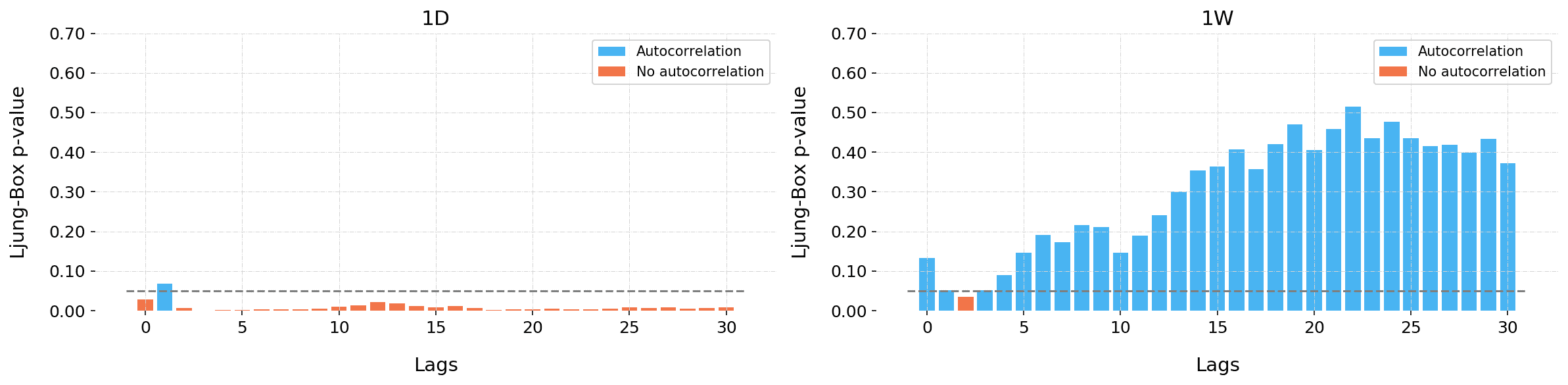}
    \caption{Ljung-Box test on ETH/BTC, from 2016-03-09 to 2019-07-01}
    \label{fig:ethbtc_lbtest}
\end{figure}

\begin{figure}[H]
    \centering
    \includegraphics[scale=0.2]{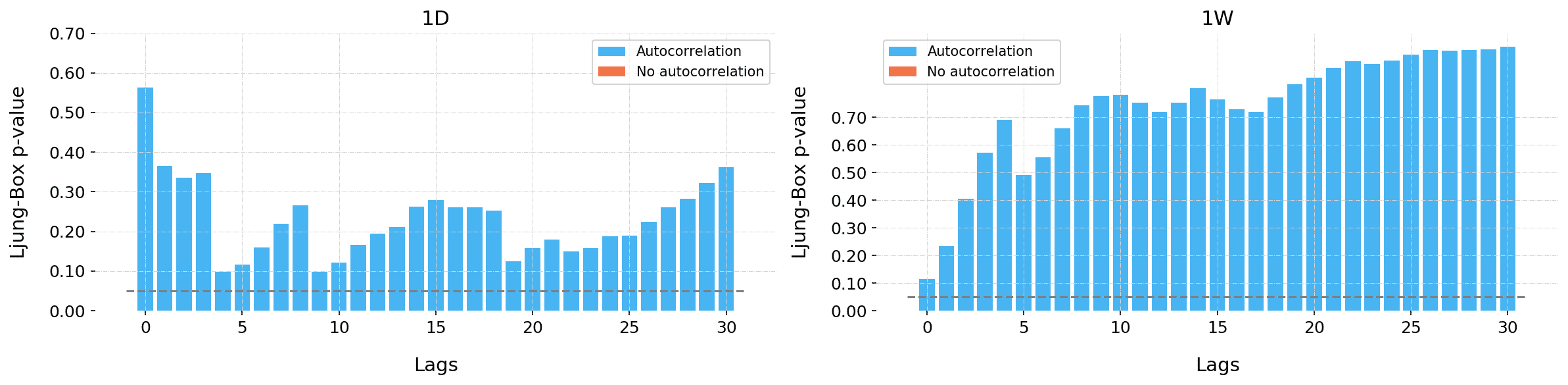}
    \caption{Ljung-Box test on XBT/USD, from 2017-10-12 to 2019-07-01}
    \label{fig:xbtusd_lbtest}
\end{figure}

In addition to Pearson's autocorrelation, Ljung-Box test's p-value was calculated for a number of lags $n = 30$ on the BTC/USD, ETH/USD, ETH/BTC markets on Bitfinex exchange and the XBT/USD market on Bitmex exchange, each on 5m, 1H, 1D and 1W time frames. For the 5m and 1H time frames for each market, p-values remained zero, $p=0$, signaling that the null hypothesis of no autocorrelation was rejected on all lags starting from the first at the threshold equal to $0.05$. Results for the 1D and 1W time frames are presented in figures \ref{fig:btcusd_lbtest} to \ref{fig:xbtusd_lbtest}. 

To reiterate, on the 5m and 1H time frames, the null hypothesis of no autocorrelation was rejected for every market on all lags, signifying that the Ljung-Box test detected the presence of autocorrelation on all four markets on lower time frames. These results are consistent with the results observed by measuring Pearson's autocorrelation coefficient. 

The picture is not as clear for the 1D and 1W time frames. For the XBT/USD pair, the p-value never dropped below the threshold on any lag. For the BTC/USD pair, the p-value was above the threshold for lower lags, dropping below the threshold for several lags and rising again later. This observation, while statistically significant, is probably too unstable to be usable. For the ETH/USD pair on the 1W time frame, the majority of lags--starting with the first--had p-value below the threshold, signifying the presence of autocorrelation. For the ETH/BTC pair on the 1D time frame, all lags except one showed p-value below the threshold.

\subsection{Rolling autocorrelation}
For the last experiment, we explored how first-order autocorrelation coefficient value changes over time. To do that, we calculated first-order autocorrelation coefficient in a rolling 1-year long window on the BTC/USD, ETH/USD, and ETH/BTC markets on Bitfinex exchange and the XBT/USD market on Bitmex exchange, each on 5m, 1H, 1D and 1W time frames.  
 
\begin{figure}[H]
    \centering
    \includegraphics[scale=0.2]{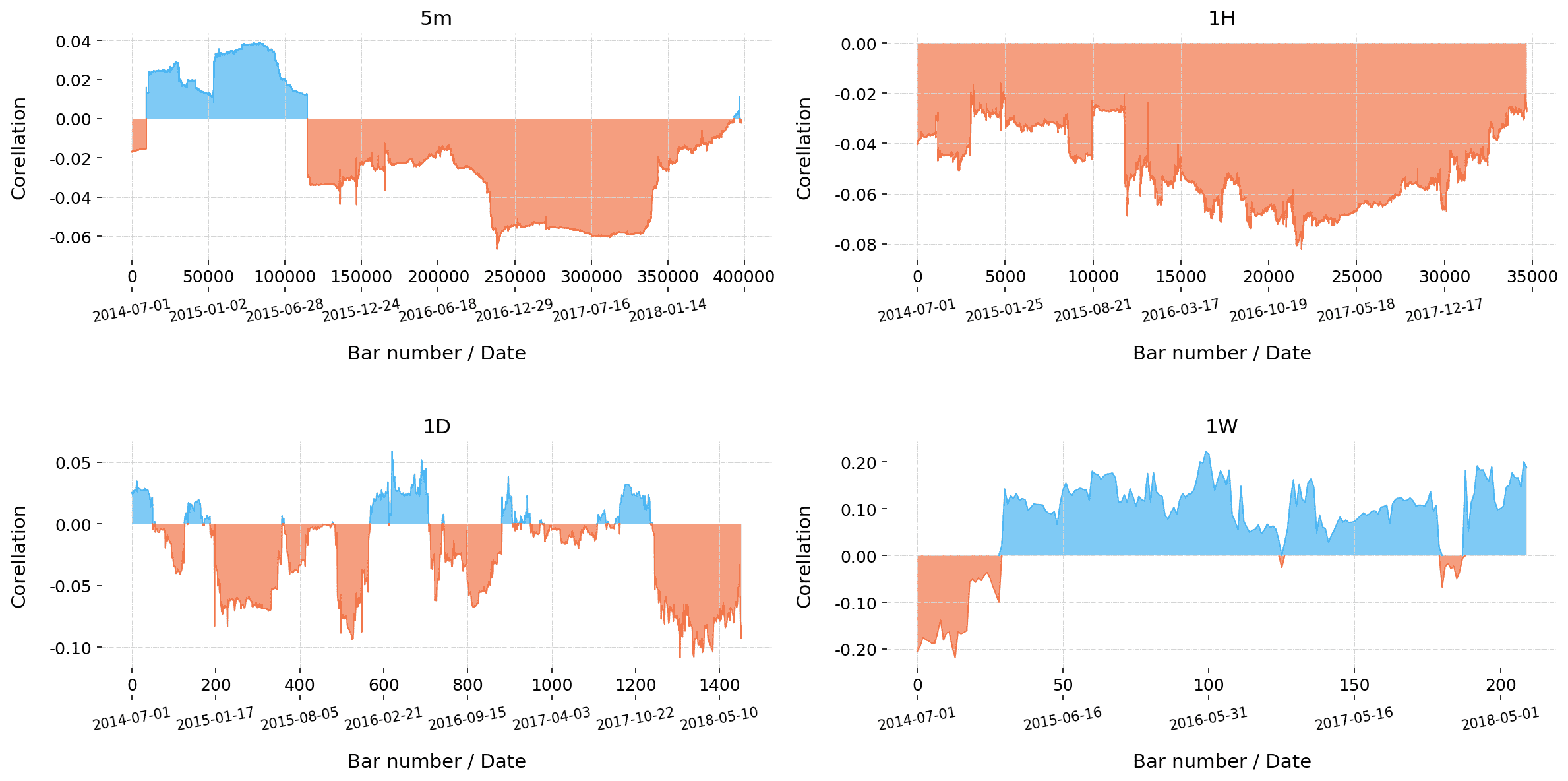}
    \caption{Rolling autocorrelation on BTC/USD, from 2014-07-01 to 2019-07-01, 1-year window}
    \label{fig:btcusd-rolling-acorr}
\end{figure}

\begin{figure}[H]
    \centering
    \includegraphics[scale=0.2]{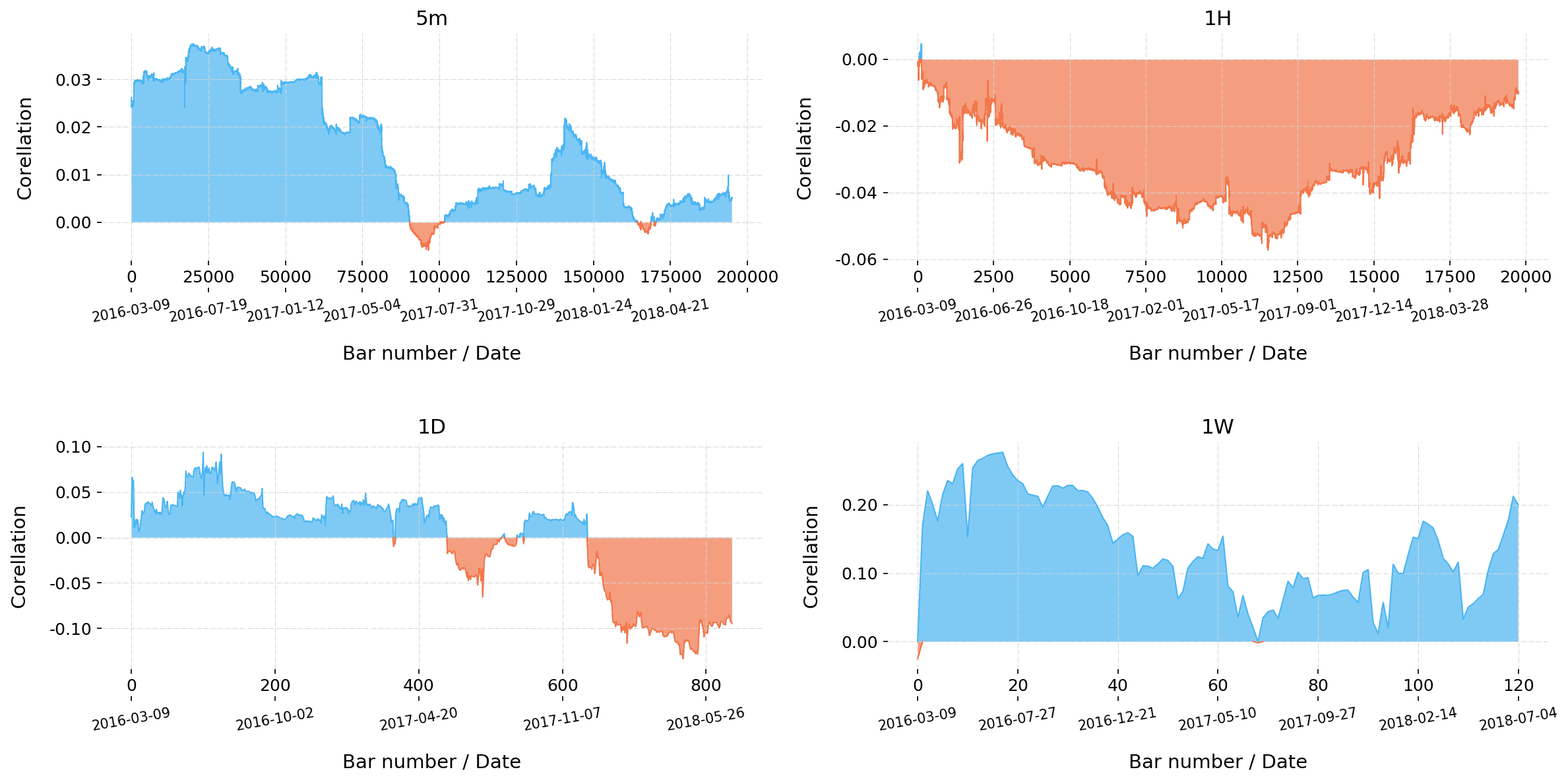}
    \caption{Rolling autocorrelation on ETH/USD, from 2016-03-09 to 2019-07-01, 1-year window}
    \label{fig:ethusd-rolling-acorr}
\end{figure}

\begin{figure}[H]
    \centering
    \includegraphics[scale=0.2]{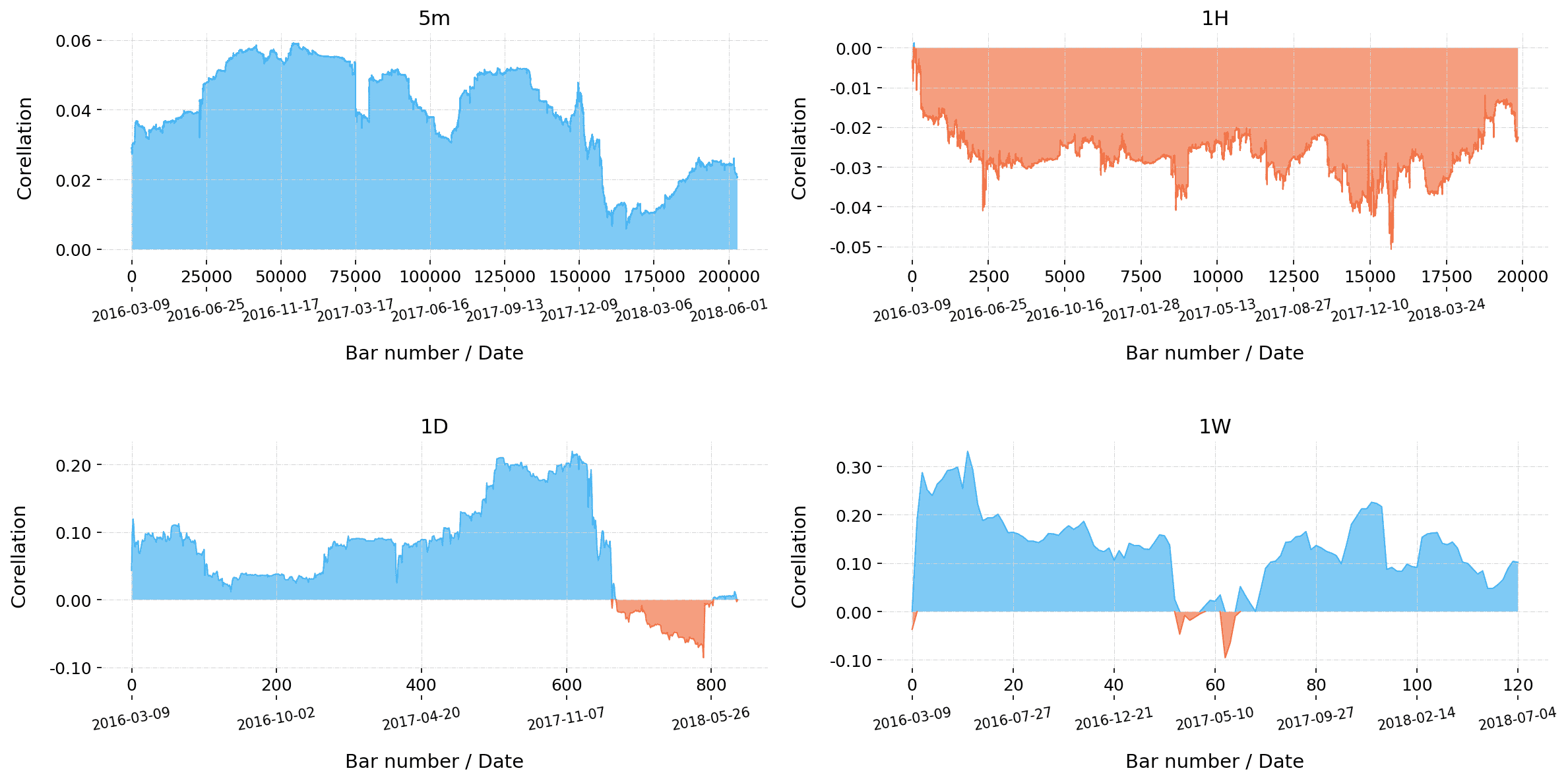}
    \caption{Rolling autocorrelation on ETH/BTC, from 2016-03-09 to 2019-07-01, 1-year window}
    \label{fig:ethbtc-rolling-acorr}
\end{figure}

\begin{figure}[H]
    \centering
    \includegraphics[scale=0.2]{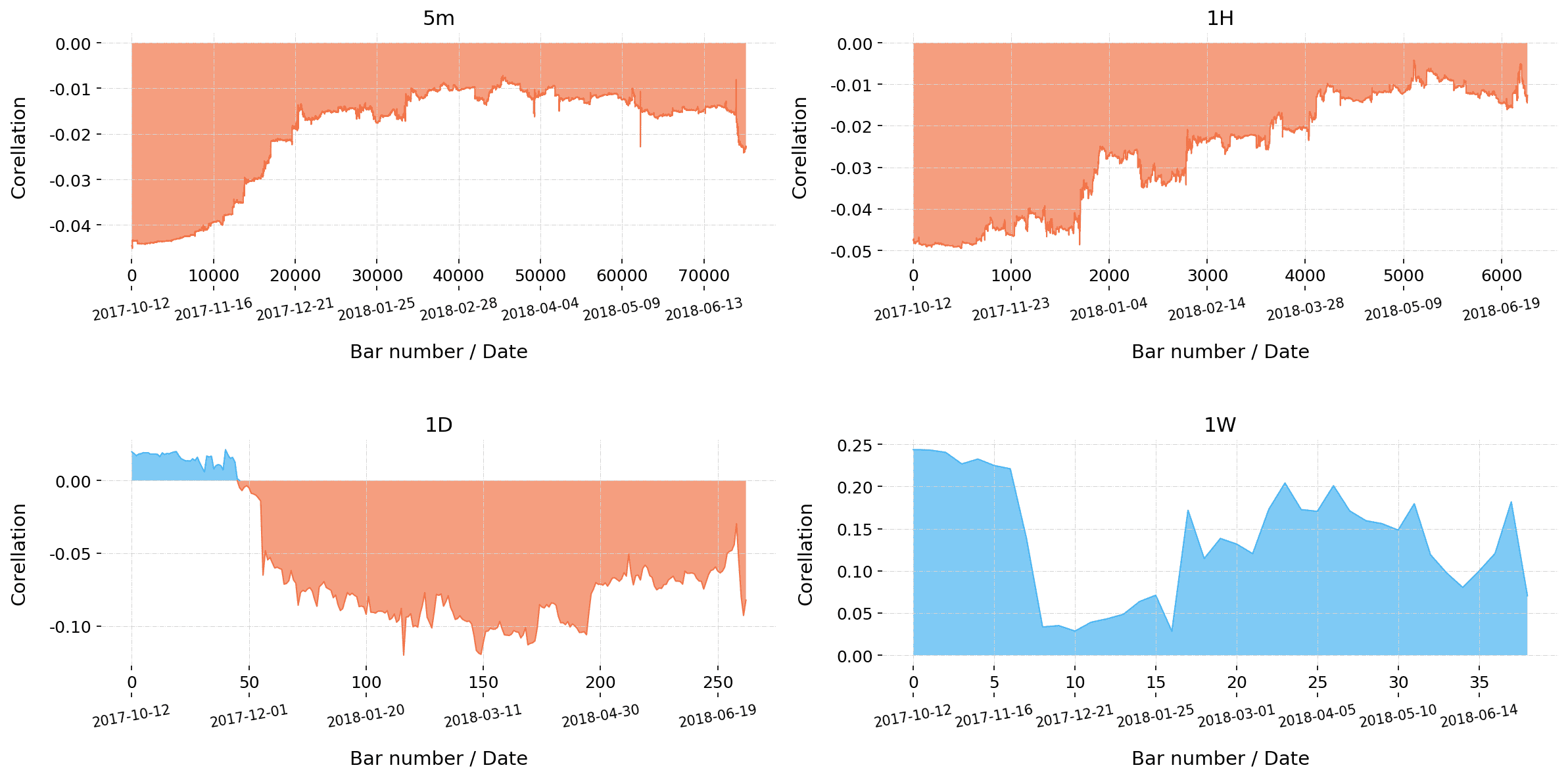}
    \caption{Rolling autocorrelation on XBT/USD, from 2017-10-12 to 2019-07-01, 1-year window}
    \label{fig:xbtusd-rolling-acorr}
\end{figure}

From the results in figures \ref{fig:btcusd-rolling-acorr} to \ref{fig:xbtusd-rolling-acorr}, it can be seen that first-order autocorrelation coefficient for the 1H time frame stays negative over time for every market. In contrast, for the 1W time frame, it stays positive most of the time in all markets. For the 5m time frame, the pictures differ between ETH and BTC pairs. Rolling autocorrelation of ETH pairs tends to stay positive over time for the 5m time frame, while for the BTC (XBT) pairs, it tends to stay negative. For the 1D time frame, rolling autocorrelation on every market behaves differently. 

\section{Discussion and conclusion}
In this paper, we explored the behavior of autocorrelation tests on the BTC/USD, ETH/USD, ETH/BTC markets on Bitfinex exchange and the XBT/USD market on Bitmex exchange, each on 5m, 1H, 1D and 1W time frames. To do so, we conducted three different experiments, measuring Pearson's autocorrelation coefficient of different orders, p-values of Ljung-Box test up to 30 lags, and first-order Pearson's autocorrelation coefficient value in a rolling 1-year window.

The initial hypotheses were as follows:
\begin{itemize}

\item There should be no statistically significant autocorrelation of any order on any time frame if the market is efficient, according to EMH.

\item The p-value of Ljung-Box test should not drop below 0.05 on any lag and any time frame if the market is efficient, according to EMH as well.

\item Rolling autocorrelation should display the same stable, close to zero results on any time frame, which follows from the first point.

\end{itemize}

According to observed results, all three hypotheses can be rejected for every market on 5m and 1H time frames. We found statistically significant autocorrelation of lower orders for all considered markets on 5m and 1H time frames by calculating Pearson's autocorrelation coefficient. The same observation was confirmed by Ljung-Box test. Rolling autocorrelation confirmed these observations to be present over time.

Results for 1D and 1W time frames are less clear and not stable and uniform enough to reject EMH or draw any practical conclusions, though rolling autocorrelation on the 1W time frame shows interesting tendency to be strongly positive on all markets.

Even though all three hypotheses were rejected and the markets were found to be inefficient according to EMH, these findings alone are insufficient to extract profits from crypto markets. However, rolling autocorrelation might constitute a useful feature to improve the predictive power of machine learning models. 

In our future work, we plan to explore whether long memory is present in the same markets as well as to measure statistical characteristics of the distribution of returns. These articles are going to be used as the baseline for further research into how we can work with financial data in ways that will present markets as less efficient and more predictable.

\bibliography{main.bib}

\end{document}